\documentclass[12pt,preprint]{aastex}

\begin{document}

\newcommand{\gsim}{\mbox{\raisebox{-1.0ex}{$~\stackrel{\textstyle >}
{\textstyle \sim}~$ }}}
\newcommand{\lsim}{\mbox{\raisebox{-1.0ex}{$~\stackrel{\textstyle <}
{\textstyle \sim}~$ }}}
\newcommand{\psim}{\mbox{\raisebox{-1.0ex}{$~\stackrel{\textstyle \propto}
{\textstyle \sim}~$ }}}
\newcommand{\vect}[1]{\mbox{\boldmath${#1}$}}
\newcommand{\lmk}{\left(}
\newcommand{\rmk}{\right)}
\newcommand{\lnk}{\left\{ }
\newcommand{\nn}{\nonumber}
\newcommand{\rnk}{\right\} }
\newcommand{\lkk}{\left[}
\newcommand{\rkk}{\right]}
\newcommand{\lla}{\left\langle}
\newcommand{\p}{\partial}
\newcommand{\rra}{\right\rangle}
\newcommand{\vex}{{\vect x}}
\newcommand{\vek}{{\vect k}}
\newcommand{\vel}{{\vect l}}
\newcommand{\vem}{{\vect m}}
\newcommand{\ven}{{\vect n}}
\newcommand{\vep}{{\vect p}}
\newcommand{\veq}{{\vect q}}
\newcommand{\veX}{{\vect X}}
\newcommand{\veV}{{\vect V}}
\newcommand{\beq}{\begin{equation}}
\newcommand{\eeq}{\end{equation}}
\newcommand{\beqa}{\begin{eqnarray}}
\newcommand{\eeqa}{\end{eqnarray}}
\newcommand{\mpc}{\rm Mpc}
\newcommand{\hmpc}{{h^{-1}\rm Mpc}}
\newcommand{\ch}{{\cal H}}
\newcommand{\lab}{\label}

\title{ Evolution of Power Spectrum in Non-Gaussian Models} 
\author{\sc Naoki Seto}
\affil{ Department of Earth and Space Science,
Osaka University,
Toyonaka 560-0043
}
\begin{abstract}
Evolution of power spectrum is studied for non-Gaussian models of
 structure formation. We generalize the dark-matter-approach to
 these models and find that the evolved spectrum at weakly nonlinear
 regime is mainly determined by a simple integral of the
 dark-matter-halo  mass function in this approach. 
  We also study the change of 
 the nonlinear spectrum  within the  current observational constraint
of the  primordial non-Gaussianity. 
\end{abstract}

\section{INTRODUCTION}
The power spectrum (or equivalently, the two-point correlation
function as its Fourier transform) is the most fundamental measure to
quantify matter 
clustering. Observational determination of it will bring us basic
information of our universe. For example we can make constraints on
cosmological parameters or on statistical aspects of the  initial matter
fluctuations that are the  seeds of structure formation. Evolution of
power 
spectrum by gravitational instability
is also an interesting problem of nonlinear physics.

It is often assumed that the initial density fluctuations obey random
Gaussian distributions. Though this is the simplest assumption from
statistical point of views, its origin is explained by the simplest
inflation scenario.  Many analyses of the large-scale structure (LSS) in
the 
universe are performed under this assumption. But other models
(including some inflationary models and defect models) predict
non-Gaussian initial 
fluctuations (Vilenkin 1985, Vachaspati 1986, Srednicki 1993, Peebles
1997, Linde \& Mukahanov 1997) and  
 possibilities of these models should be observationally investigated and
have recently called much attentions ({\it e.g.} Peebles 1999a, 1999b,
Ferreira,
G\'orski \& Magueijo
1998, Robinson, Gawiser \& Silk 1998, 2000, Koyama, Soda \& Taruya 1999,
Willick 2000). These recent analyses involve the large-scale structure,
properties ({\it e.g.} time evolution, spatial clustering) of clusters
and CMB anisotropies. 
It now becomes clear that defect models  cannot generate the observed
CMB data by their own account
 but inflation models can explain them well ({\it e.g.} Albrecht 2000).
Recent CMB data  
  by BOOMERanG (Lange et al. 2000)
and MAXIMA (Balbi et al. 2000) seem to suggest a weaker secondary peak
than 
expected in a simple inflation scenario. Bouchet et al. (2000) discussed
that this might be resolved by a hybrid scenario (inflation+cosmic
string) in which some component of perturbations come from topological
 defects.

 As for the quantitative measurement of the primordial non-Gaussianity,
 Contaldi et al. (2000) studied   the
third-order moment of the temperature anisotropies using 4yr COBE
Differential Microwave Radiometer (DMR) data  and found their most
conservative estimate as $\lla(\Delta T)^3\rra=-6.5\pm 8.7  [10^4 \mu{\rm
K}^3]$ (95\% CL). Even though the magnitude of the error bar is large 
due to the limited sensitivity and angular resolution, 
traditional Gaussian 
model ($\lla(\Delta T)^3\rra=0$) is consistent with their result.
Feldman et al. (2000) analyzed 
the bispectrum of the IRAS PSCz catalog (Saunders et al. 2000) and
obtained a constraint for 
the primordial dimensionless skewness as
$S\equiv \lla\delta^2\rra/\lla\delta^2\rra^{3/2}<0.52$ (95\% CL) for
$\chi_N^2$-models.  Physics of CMB are much simpler than that of the
galaxy distribution. The signal of the primordial non-Gaussianity
could be masked by effects of biasing or nonlinear gravitational
evolution (Verde et al. 2000).  Observed data of both CMB and LSS will
be significantly 
improved 
from  MAP and Plank satellites,  SDSS and  2dF surveys. Verde et
al. (2000) insisted that CMB  would be a better  probe of the primordial
non-Gaussianity for several non-Gaussian
 models and LSS data would be useful to
constrain the biasing if primordial Gaussianity is suggested by CMB map.

So far nonlinear effects of matter clustering from non-Gaussian initial
conditions are studies by numerical simulations
 (White 1999, Robinson \& Baker
2000 and references therein) or by perturbative methods ({\it e.g.} 
Fry \& Scherrer
1994, Chodorowski \& Bouchet 1996, Scoccimarro 2000, Durrer et al. 2000).
In this article we study evolution of the matter
 power spectrum for non-Gaussian
models  using the dark-matter-halo approach that
can be traced back to Peebles (1974) and McClelland and Silk (1977) (see
also Scherrer \& Bertschinger 1991). For
Gaussian models this approach is confirmed to reproduce numerical
results very well from linear to nonlinear scales and regarded as an
excellent tool to explore the  matter clustering. It seems interesting
to see how the nonlinear spectrum can be different within the currently
allowed region of the primordial non-Gaussianity.

Many works have been recently
performed with this approach for Gaussian models. For example nonlinear
evolution of the power spectrum or the
 bispectrum of density field is studies by
 Ma 
\& Fry (2000a, 2000b, 2000c), Seljak (2000), White (2000), Peacock \&
Smith (2000),  
Scoccimarro et al. (2000) and weak lensing effects  by Cooray, Hu \&
Miralda-Escude (2000) and  Cooray \& Hu 
(2000). Komatsu \& Kitayama (1999) investigated the Sunyaev \& Zeldovich
effect using a similar method.
Here we simply extend this approach to non-Gaussian models.  This
approach is made from three basic ingredients (i) the mass function of
dark-matter-halos, (ii) the bias parameter of halos relative to the 
large-scale density fluctuations and (iii) the density profile of halos.
Numerical analysis (Robinson \& Baker 2000) supports that a
straightforward 
extension of the Press \& Schechter (1974) mass function 
 is effective for 
 non-Gaussian models.  But validity of  simple extensions of
two other elements have not been checked numerically. Therefore we make a
careful analysis  and limit our investigation
 in the range where  results are
expected not to depend largely on the details of these two elements.  
  Our goals in this article  are set to the following
two 
points.  Firstly, we intend to roughly understand the effects of
primordial non-Gaussianity on the weakly nonlinear evolution of the
power spectrum. Secondary, we estimate the difference of the evolved
spectra within the current observational  constraint of the primordial
non-Gaussianity. 

This article is organized as follows. In \S 2 we describe basic
properties of the dark-matter-halo approach with 
its basic three ingredients, namely, the mass function of halos (\S
2.1), the 
bias parameter (\S 2.2) and the density profiles of halos (\S 2.3). In \S
3  we analyze toy models to clarify the general effects of
non-Gaussianity on the weakly nonlinear evolution of power spectrum. In
\S 4 we investigate realistic models. We study a flat
$\Lambda$-dominated model ($\Omega_0=0.3, \lambda_0=0.7$ and $h=0.7$) 
with a fixed linear CDM spectrum and evaluate evolution of power spectra
for log-normal probability distribution functions that are
 motivated by work of Robinson \& Baker (2000). \S
5 is devoted to a brief summary. In appendix A we discuss the scale
dependence of the skewness parameter for $\chi^2_N$-models.
\section{DARK MATTER HALO APPROACH}

In the dark-matter-halo approach the nonlinear density field is given by
superposition of halos with various masses. The density
profile of each halo is assumed to be  determined by its mass.
The nonlinear power spectrum  $P_{NL}(k)$ is  constituted
by two terms as
\beq
P_{NL}(k)=P_{2h}(k)+P_{1h}(k).\lab{tot}
\eeq
The two-halo term $P_{2h}(k)$ counts contributions of two points coming
from two different halos and is given  by
\beq
P_{2h}(k)\equiv \lkk \rho^{-1}_b\int dm N(m) m b(m) u(k,m) \rkk ^2 P_{L}(k),\lab{2h}
\eeq
where $P_{L}(k)$ is the linear power spectrum.  The one-halo term
$P_{1h}(k)$ counts two particles within the  same halo
\beq
P_{1h}(k)\equiv\rho^{-2}_b \int dm N(m) m^2 u(k,m)^2.\lab{1h}
\eeq
In the above  equations $\rho_b$ is the background density of the
universe  and
$N(m)dm$ is the number density (mass
function) of halos with 
mass 
from 
$m$ to $m+dm$.  The factor 
$b(m)$ is the   bias parameter and represents distribution of halos
relative to the 
large-scale density fluctuations. The function $u(k,m)$ is the
Fourier transform of the density profile of a dark-matter-halo with mass
$m$. The original form of the 
dark-matter halo approach was  proposed by Peebles (1974). This approach
 has recently called much attention as our
understandings of its ingredients have largely developed. 
In the following three subsections we study these basic ingredients.

\subsection{Mass Function}
We use the Press \& Schechter (1974) formalism for the
 mass function $N(m)$ of
collapsed objects. This formalism
 was first  given for Gaussian random  fields
but can be straightforwardly extended to general models ({\it e.g.}
Lucchin \& Matarrese 1989, Chiu,
Ostriker \& 
Strauss 1998, Koyama, Soda \& Taruya 1999, Robinson, Gawsier \& Silk
2000, Willick 2000, Matarrese, Verde \& Jimenez 2000).  For simplicities
we assume that the  linear 
 one-point probability distribution function (hereafter PDF)
 $p(\delta,\sigma)$
for the smoothed 
density field  is written as
\beq
p(\delta,\sigma)d\delta=\sigma^{-1}p(\delta/\sigma)d\delta,
\eeq
which means that the shape of the PDF is scale invariant.
This
ansatz is correct for the 
random Gaussian  fluctuations, but not guaranteed
for general non-Gaussian models. We comment on this scale dependence in
\S4 and appendix A.  

The volume fraction of points that belong to collapsed objects with mass
larger than $m$ is given as follows
\beq
F^{-1}\int^\infty_{\delta_c/\sigma(m)}p(\nu)d\nu,\lab{vf}
\eeq
where $\delta_c\simeq3(12\pi)^{2/3}/20=1.69$ is the critical linear
density 
contrast for the 
spherical top-hat collapse  and
$\sigma(m)$ is the root-man-square mass fluctuation in a sphere
of mass scale $m$.  $F$ is the
normalization factor to 
account for all  masses in the universe 
\beq
F=\int^\infty_0p(\nu)d\nu.\lab{norm}
\eeq
For a symmetric profile $p(-\nu)=p(\nu)$ we have $F=1/2$ (see also Bond et
al. 1991, Lacey \& Cole 1993).
From equation (\ref{vf}) the number density of halos with mass from $m$ to
$m+dm$ is given as 
\beq
N(m)dm = -F^{-1} \rho_bm^{-1} \frac{\p}{\p m}\int^\infty_{\delta_c/\sigma(m)}p(\nu)d\nu = -F^{-1} \rho_b p(\nu)m^{-1} \nu \frac{d\ln\sigma(m)}{dm},\lab{mf}
\eeq
where we have denoted the normalized density contrast
$\nu\equiv \delta_c/\sigma(m)$. We define the 
nonlinear mass $m_*$  by equation
 $\sigma(m_*)=\delta_c$. For power-law models
with 
$P_L(k)\propto k^n$ the linear mass fluctuation $\sigma(m)$ is given as
\beq
\sigma(m)= \delta_c \lmk\frac{m}{m_*} \rmk^{-(n+3)/6} .
\eeq

In this subsection we have followed the  basic analysis of the Press \&
Schechter method. For random Gaussian fluctuations  some
refinements have been proposed to reproduce  N-body data better
({\it e.g.} Sheth \& Tormen 1999, Jenkins et al. 2000). 
If we write down the mass function in the following form
\beq
N(m)=-A \sqrt{\frac2\pi \alpha}\frac{d\ln \sigma}{d\ln m}\frac{\rho_b}{m^2}
\nu e^{-\alpha \nu^2/2}(1+(\alpha \nu^2)^{-p}).\lab{mst}
\eeq
The original Press \& Schechter function corresponds to $(A, p,
\alpha)=(0.5,0,1)$. The mass function given by Sheth \& Tormen is given
by parameters  $(A, p, \alpha)=(0.322,0.3,0.707)$.

Robinson \& Baker (2000) have  recently
confirmed that the simple extension of the
 Press \& Schecheter formula (\ref{mf}) can accurately
 fit the mass function of clusters for several non-Gaussian models with
typical error $\sim25$  percent (see also Avelino \& Viana 2000).  This result
encourages our analytical study.

\subsection{Bias Parameter}
Next we study the bias parameter $b(m)$ using (i) the peak-background
splitting 
method (Kaiser 1984) and (ii) a simple extension of the Press \& Schechter
method  (Mo \& White 1997, Mo, Jing \& White 1997, Sheth \& Lemson 1999,
Sheth \& Tormen 1999).  When a large-scale perturbation $\delta_{LS}$ is
added to the
local 
(small scale) density field $\delta_{SS}(\vex)$, the  density
contrast for $\delta_{SS}(\vex)$ that is 
critical to the  spherical top-hat collapse
becomes $\delta_{c}-\delta_{LS}$. Then the Eulerian bias parameter for
halos with mass $m$  becomes  (Mo \& White 1997, Robinson,
Gawiser \& Silk 1998, Koyama, Soda \& Taruya 1999)
\beq
b(m)=1-\frac{d\ln
N(m)}{d\delta_c}=1-\frac1{\delta_c}\lmk\frac{p'(\nu)}{p(\nu)}\nu  +1\rmk.\lab{bias}
\eeq
From equations (\ref{mf}) and (\ref{bias}) we can easily confirm the following equation
\footnote{This equation holds for general non-scaling PDF in the form
$p(\delta,\sigma)$.} 
\beq
\rho_b^{-1}\int_0^\infty b(m)N(m)m dm=1.\lab{lar}
\eeq
As we see in \S 4 this equation is useful to know the  behavior of
the nonlinear power spectrum $P_{NL}(k)$ at large scale (small $k$).
There are several analytical works about the bias parameter for
non-Gaussian models (Robinson,
Gawiser \& Silk 1998, Koyama, Soda \& Taruya 1999). But validity of the
formulas 
 (\ref{bias}) for a non-Gaussian PDF have not been checked 
with numerical simulations. But we expect that it works well, as this
formula    is simply derived from the Press \& Schechter method that has
been
checked numerically for some non-Gaussian models
 as commented in the previous subsection.

For the mass function (\ref{mst}) given for random Gaussian fluctuations
we have
\beq
b(\nu)=1+\frac{\alpha\nu^2-1}{\delta_c}+\frac{2p}{\delta_c(1+(\alpha
\nu^2)^{p})}.\lab{stb}
\eeq

\subsection{Density Profile of Dark Matter Halo}

Recent numerical simulations suggest that the average density profile of
cold-dark-mater (CDM)
 halos has nearly universal shape $\rho_m(r)$ characterized by its
mass $m$ (Navarro, Frenk \& White 1996, 1997, Moore et al. 1998, 1999,
Kravsov et al. 1998,
Fukushige \& Makino 1998, Jing \& Suto 2000). These numerical
 simulations are performed with Gaussian initial conditions,  but here
we assume that the density profile depends  only weakly
 on the linear PDF as
the halo profile would strongly reflect nonlinear gravitational dynamics
rather than the initial conditions. 
This assumption may be
 a potential weakness of our dark-matter-halo approach 
  especially at
strongly nonlinear regimes. However as we see in \S 4, the power spectrum
$P_{NL}(k)$ at weakly nonlinear regimes does not depend details of the
halo profile. 
In this article we adopt the universal
 profile of dark matter
halo 
given in Navarro, Frenk \& White (1997) as follows
\beqa
u_m(r)&\equiv& \frac{\rho_m(r)}m\lab{prod}\\
&=&\frac{fc^3}{4\pi
R_{vir}^3}\frac1{(cr/R_{vir})^{-\alpha}
(1+cr/R_{vir})^{3+\alpha}},\lab{pro} 
\eeqa
where $R_{vir}$ is the virial radius of the halo. The halo mass $m$ is
written by  $R_{vir}$ as $m=4\pi R^3_{vir}\rho_b\Delta/3$ with the
 average
density contrast 
$\Delta=200$ (Einstein de-Sitter model) and $\Delta=340$ (flat model
with $\Omega_0=0.3$ and $\lambda_0=0.7$). The concentration parameter $c$
determines the transition of 
inner and outer regions of a halo and depends on its mass $m$, roughly
reflecting the 
formation epoch of the  halo (Ma \& Fry 2000b). We have $c=\beta(n)
\sigma(m)$ with a coefficient $\beta(n)$ that depends on the linear  power
spectrum.  
The normalization factor $f$ in equation (\ref{pro}) is given by 
\beq
f=(\ln (1+c)-c/(1+c))^{-1}.
\eeq
The density profile (\ref{pro}) in central region becomes $\rho(r)\propto
r^\alpha$. The original result in Navarro, Frenk \& White (1996) was
$\alpha=-1$. But some other simulations show different results. For
example, Moore et al. (1998, 1999) found $\alpha=-1.5$ and Jing \& Suto
(2000) argued that the index $\alpha$ is not a universal value
 but depends on
the mass scale of halos. As we see later, the nonlinear evolution of
power spectrum in weakly nonlinear regime is not sensitive to the
details of the halo profile (see also Seljak 2000). Thus we use
$\alpha=-1$ in this article.

Finally  the Fourier transformed function $u(k,m)$ is defined in terms
of the real space density profile $u_m(r)$ as
\beqa
u(k,m)&=&\int d^3 x \exp(-i\vex\cdot\vek)u_m(r)\\
&=&4\pi \int \frac{\sin kr}{kr}r^2 u_m(r)dr. 
\eeqa
For the profile (\ref{pro}) an explicit analytic formula of $u(k,m)$ was given by
Scoccimarro et al. (2000). At small wave number $k$ the above Fourier
transformation becomes a simple volume integral and we have $u(0,m)=1$
due to the definition of $u_m(r)$ given in equation (\ref{prod}).

\section{TOY MODELS}

In this section we discuss effects of initial non-Gaussianity on the
evolution of the power spectrum in Einstein de-Sitter background. We
intend to extract out their general 
relations.   
From this standpoint we do not try to mimic realistic PDFs
so seriously but
study toy models that are different from Gaussian distribution. 
We also limit our analysis to scale-free initial spectra.
With results obtained in this section we study  models relevant for
observational cosmology in \S 4.
\subsection{Density Distribution Functions}

 We use the following four PDFs for the
 linear smoothed density
fields.  As commented in \S2.1 we assume that the PDFs are
scale invariant. The first example is the standard
Gaussian model (G-model),
\beq
p_G(\nu)=\frac1{\sqrt{2\pi}}\exp\lmk-\frac{\nu^2}2
\rmk~~~-\infty<\nu<\infty .
\eeq
This PDF is symmetric around the mean density $\nu=0$ and we have $F=0.5$.
Here the factor $F$ is defined in equation (\ref{norm}).

The second example is the Chi-square ($\chi^2$)-model (C-model) with two
degrees of 
freedom. A $\chi^2$-model is often adopted as a typical non-Gaussian
model and its origin in cosmological context  was recently proposed by
Peebles (1999a) using an inflation model. The PDF for the  unsmoothed
field 
is
\beq
p_C(\nu) =\cases{
\exp(-\nu-1) &$ (-1<\nu<\infty),$ \cr
0 &$ (\nu \le -1),$ \cr 
}
\eeq
with $F=1/e$. We should notice that the PDF for the  smoothed density
field  is  same as the above unsmoothed one.
This point is discussed in the next section and appendix A.

Next example is a log-normal distribution. This distribution is
characterized by a parameter $\alpha(\ge 0)$ as follows
\beq
p_{LN}(\nu,\alpha) =\cases{ \displaystyle
\frac{\alpha}{(1+\nu\alpha) \sqrt{2\pi
\ln(1+\alpha^2)}}\exp\lkk-\frac{\lnk\ln(\sqrt{1+\alpha^2} 
(1+\nu\alpha)) \rnk^2 }{2\ln(1+\alpha^2)} \rkk & $-\alpha^{-1}<\nu<\infty$, \cr
0 &$ \nu \le -\alpha^{-1}$. \cr 
}\lab{lnd}
\eeq
At $\alpha=0$ we recover the Gaussian distribution
$p_G(\nu)=p_{LN}(\nu,0)$. The normalization factor $F$ for this
distribution is given in terms of the error function as
\beq
F=\frac1{2}{\rm erfc}\lkk \frac12 \sqrt{\frac{\ln(1+\alpha^2)}2} \rkk,
\eeq
where  the error function is defined as ${\rm
erfc}(x)=\sqrt{2/\pi}\int_x^\infty \exp(-y^2/2)dy$.

  Here we add two models generated  from
equation (\ref{lnd}).  We define the LN1-model by
\beq
p_{LN1}(\nu)=p_{LN}(\nu,1)
\eeq
This profile has  larger positive
 tail than the G-model. We have $F=0.3386$.
The final example (LN2-model) is obtained by a simple replacement
of the previous model as  
\beq
p_{LN2}(\nu)\equiv
p_{LN1}(-\nu)=p_{LN}(-\nu,1).
\eeq 
Note that we have $p_{LN2}(\nu)=0$ for $\nu>1$. Existence of halos with
mass larger than $m_*$ is prohibited in this model.
As we see in the following section, the shape of the positive tail
is very important to study nonlinear evolution of the  power spectrum. The
probability for $\nu\ge 3$ is $1.34\times 10^{-3}$ (G-model),
$1.83\times 10^{-2}$ (C-model), $1.87\times 10^{-2}$ (LN1-model) and
0 (LN2-model).  For $\nu \ge 5$ we have
$2.87\times 10^{-7}$ (G-model),
$2.48\times 10^{-3}$ (C-model), $5.11\times 10^{-3}$ (LN1-model) and
0 (LN2-model).
We can easily confirm that all of these models have vanishing means
\beq
\lla \nu \rra=\int_{-\infty}^\infty  \nu p(\nu)d\nu=0,
\eeq
and are normalized properly
\beq
\int_{-\infty}^\infty  p(\nu)d\nu=1,~~~
\lla\nu^2\rra=\int_{-\infty}^\infty \nu^2 
p(\nu)d\nu=1. 
\eeq

The skwness parameter  $S\equiv
\lla\nu^3\rra/\lla\nu^2\rra^{3/2}$ is a fundamental measure to
characterize PDFs. \footnote{Definition of   skewness parameter
 is different
from that usually  used  for Gaussian initial condition
(Bouchet et al. 1992).  } We have $S=0$ (G-model), 2 (C-model), 4 (LN1-model)
and $-4$ (LN2-model). For the general log-normal distribution
$p_{LN}(\nu,\alpha)$ we have 
\beq
S_{LN}=\alpha (3+\alpha^2).\lab{lns}
\eeq  
We use this relation in section 4.
\subsection{Results}

Using ingredients described in \S 2 and various linear PDFs given in \S 3
we calculate the nonlinear power spectrum $P_{NL}(k)$ with equations (\ref{tot}),
(\ref{2h}) and (\ref{1h}).
We use power-law models for initial  spectra
\beq
P_{L}(k)=Ak^n,
\eeq
where $A$ is the normalization factor but irrelevant for present analysis.
 Time evolution of power spectrum
can be reduced to rescaling of spatial length due to our scale-free
initial conditions.   We  introduce the nonlinear
wave-number $k_{NL}$ that is defined by
\beq
\frac1{2\pi^2}\int_0^{k_{NL}}P_L(k)k^2dk=1.\lab{knl}
\eeq
This wave-number is the most fundamental scale for the present
analysis. In this article  we call  the weakly nonlinear scale for the
wave number $k$ 
\beq
0<k\le k_{NL}. 
\eeq

We first calculate the nonlinear power spectra $P_{NL}(k)$ for random
Gaussian fields 
and compare our analytical predictions  with fitting formula that is
 obtained from
N-body simulations with the same Gaussian initial conditions
 (Peacock \& Dodds 1996, see also Hamilton et
al. 1991, Jain, Mo \& White 1995). We fix the coefficient $\beta(n)=3$ for
the concentration parameter $c(m)$, but our results are insensitive to
this parameter.  It is found that our model cannot
reproduce the numerical results for spectral index with $n=0$ in
contrast to 
the case with $n\lsim -1$. We can see this tendency from figures 1.a to
1.c. 
It is not clear why this dependence  appears. Note that the evolved
spectrum (thick solid line) for $n=-1$ obtained from the fitting formula
is smaller than the linear spectrum (dash-dotted line) as predicted by
the second-order perturbation theory ({\it e.g.} Makino, Sasaki \& Suto
1992). In the framework of the dark-matter-halo approach this fact might
be 
related to the possibility that the total halo mass fraction does not
converge to unity (Jenkins et al. 2000) and  the constraint (11) does not
hold.  As we see below, without the constraint (11) the two-halo term $P_{2h}(k)$ can be
smaller than the linear spectrum $P_{L}(k)$
 with nonnegligible amount  at weakly nonlinear scale.

 Here we limit
our analysis for spectral indexes with $n\le -1$ where our approach is
expected to work well. These are close to the effective slope of a typical
CDM power spectrum at weakly nonlinear regime $\sigma\lsim 1$.
In figures 1.a to 1.c we plot the nonlinear power spectrum $P_{NL}(k)$
in the form  $\Delta(k)$ defined by
\beq
\Delta(k)\equiv \frac{P_{NL}(k)k^3}{2\pi^2},
\eeq
for spectral indexes $n=-2,-1.5$ and $-1$. In these figures 
contributions from the one-halo term $P_{1h}(k)$ and the two-halo term
$P_{2h}(k)$ 
are presented 
separately. Note that our analytical prediction for the Gaussian models
reproduce the  results from N-body simulations
  well in the range $k\le k_{NL}$ for $n\le -1.5$ This fact
indicates validity of the dark-matter-halo approach.
For $n=-2$ model we also evaluate the nonlinear spectrum with using
generalized formulas (\ref{mst}) (\ref{stb}) as in Seljak (2000).
Our results shows good agreement with this results. 
in figures 1.a to 1.c w have used the fitting formula of Peacock \&
Dodds (1996).  We find that the fitting formula of Jain et al. (1995)
agrees with our analytical result better for $n=-2$ model but worse for
$n=-1$ model.

Next we discuss behavior of the nonlinear power spectrum  $P_{NL}(k)$
for various PDFs given in \S3.1. 
We present numerical results in figures 1.a to 1.c. These are obtained
by fully using the dark-matter-halo approach. In this calculation we
evaluate all the three elements, the mass function $N(M)$, the bias
parameter $b(M)$ and the halo profile $u(k,M)$. But information of the
mass function is most important at weakly nonlinear regime, as we see
below.

 For  all models  the two-halo term in
the range $k\ \lsim  k_{NL}$ is almost same as the linear power
spectrum $P_{L}(k)$, as shown on figures for the two-halo term.  In
this regime we have $u(k,m)\simeq 1$ 
in the dominant contribution of the  integral in equation 
(\ref{2h}) (see also figure 2).  Then we obtain the following relation  from equation (\ref{lar})
\beq
\int_0^\infty dm N(m)m b(m)u(k,m)\simeq \int_0^\infty  dm N(m)m
b(m)=\rho_b,
\eeq
and
\beq
P_{2h}(k)\simeq P_L(k),\lab{2hs}
\eeq
(see also Sheth et al. (2000) for a same kind of analysis for the
two-point correlation function).

Therefore we can roughly understand the weakly nonlinear evolution of
power 
spectrum only by 
analyzing  the one-halo term 
\beq
P_{1h}(k)=\rho_b^{-2}\int m^2 N(m)u(k,m)^2 dm.\lab{1hs}
\eeq 

As shown in figures 1.a to 1.c, the contribution of $\Delta(k)$ from
one-halo term $P_{1h}(k)k^3/(2\pi^2)$ is nearly on a straight line
proportional to $k^3$. This means $P_{1h}(k)\simeq const$ and indicates
that $k$-dependence caused by the halo profile $u(k,m)$ is weak.
  Thus we approximate the one-halo term at weakly nonlinear
regimes $k\lsim k_{NL}$ by setting $u(k,m)=1$
\beq
P_{1h}(k)\simeq P_{1h}(0)=\rho_b^{-2}\int dm N(m)m^2.\lab{1h0}
\eeq

We have a constraint for the function $mN(m)$ due to the conservation of
mass as
\beq
\int_0^\infty dm N(m)m=\rho_b,
\eeq
and $P_{1h}(0)$ in equation(\ref{1h0}) becomes larger for a mass function
$N(m)$ with more 
abundant massive objects. Thus we can roughly understand the weakly
nonlinear correction of power spectrum using information of mass
function in the form given in $P_{1h}(0)$.

Note that the present analysis at weakly nonlinear regime
$k\le k_{NL}$ based on equations (\ref{2hs}) and
(\ref{1hs}) does not depend on details of the halo density profile.
Therefore in this regime we should not be troubled seriously with
potential problems about the reliability of the universal profile for
non-Gaussian models.  This also means that our analysis is not sensitive
to  the bias
parameter $b(m)$ (eq.[\ref{bias}]).
In the region where $u(k,m)=1$ is a good approximation, we need the bias
parameter only in the form of equation (\ref{lar}). But this equation
represent a 
simple requirement that the  large-scale fluctuation is analyzed
perturbatively $P_{NL}(k)\simeq P_L(k)$.
Here we should notice that we have not taken into account the nonlinear
biasing  effects or  the exclusion effects of halo-halo
clustering. These effects might be important in weakly nonlinear regime
(Scoccimarro et al. 2000), but  we neglect them in this article. At
present there is no simple prescription to deal with these two opposite
effects. This is a generic problem of the dark-matter-halo approach.

Using results for $n=-2$ model we  discuss effects of the linear
PDF on the weakly 
nonlinear evolution of the 
 power spectrum.  As shown in figure 1.a the nonlinear
correction for the LN2-model is very weak. In this model there are no
halos with $m\ge m_*$ (see figure 2) and the integral $\int N(m)m^2 dm $
is much smaller than other models. The weakly nonlinear effect is
strongest for the LN1-model as expected from its
 mass function in figure 2.
Even at $k\simeq k_{NL}$ magnitude of the nonlinear power spectra can
differ by a factor of 10 depending on the initial PDFs.

Next we compare the behavior of nonlinear power spectrum
 $P_{NL}(k)$ for different spectral indexes $n$. 
As $n$ becomes larger, the nonlinear correction becomes smaller (figures
1a-1c). All 
nonlinear power spectra  $P_{NL}(k)$ for $n=-1$ model
 are very close to the linear spectrum $P_L(k)$, 
but these for $n=-2$ model depend largely on the adopted PDF. This 
 $n$-dependence can be also understood by using the integral $P_{1h}(0)$
given in equation (\ref{1h0}).
We have the following simple
 relation for the  root-mean-square mass fluctuations
$\sigma(m)$ 
\beq
\frac{\delta_c}{\sigma(m)}=\nu=\lmk \frac{m}{m_*} \rmk^{(n+3)/6}.
\eeq
As the index $n$ increases,  abundance of large  mass objects
 becomes smaller. Then the magnitude of the
one-halo term
 $\simeq P_{1h}(0)$
decreases and  we have smaller  nonlinear corrections.

\section{CDM SPECTRUM}

In this section we study nonlinear evolution of the power spectrum for
 realistic models.  We fix the cosmological parameters at $h=0.7$,
$\Omega_0=0.3$ and $\lambda_0=0.7$ that are compatible with
 CMB ({\it e.g.}
Lange et al. 2000) and
high-redshift  SNe
data (Riess et al. 1998,  Perlmutter et al. 1999). We use the
primordially scale  
invariant ($n=1$) spectrum with CDM transfer function from Bardeen et
al. (1986). The power spectrum is COBE-normalized and we have
$\sigma_8\simeq0.9$. For these parameters the nonlinear wavenumber
$k_{NL}$ defined in equation (\ref{knl}) becomes $k_{NL}=0.19$Mpc.

Robinson \& Baker (2000) studied evolution of the cluster abundance in
 representative non-Gaussian 
models, such as, Peebles isocurvature cold-dark-matter
 (ICDM) model or cosmic string model. 
This ICDM model corresponds to a $\chi^2$-model with one degree of
freedom (see appendix A). Peebles (1999b) show that the power spectrum for
this model with  parameters $\Omega_0=0.2$, $\lambda=0.8$ and
$n_\phi=2.4$ (see eq. [\ref{nphi}] for definition of $n_\phi$) is in
good agreement with observed spectrum of CMB and the large-scale structure.

Robinson \& Baker (2000) also analyzed PDFs of smoothed linear density
fields for these non-Gaussian
 models and found that the PDFs are well fitted by the
log-normal distributions  that are nearly scale independent at weakly
nonlinear scale. For example, they did not detect scale dependence for the
ICDM model and the string+HDM model (their table 1).  Thus, motivated by
their work,  we
investigate the  log-normal PDFs given in equation 
(\ref{lnd}) and neglect the scale dependence. We characterize the 
non-Gaussianity of 
a log-normal PDF by the skewness parameter $S\equiv
\lla\delta^2\rra/\lla\delta^2\rra^{3/2}$ (see eq.[\ref{lns}]). 
For  models in Robinson \& Baker (2000) we can estimate the
parameter $S$ from their table 1. We have $S\simeq 2.2$ (somewhat
smaller than result  $\simeq 2.5$ given in Peebles 1999b) for the ICDM
model with $n_\phi=-2.4$ and $S\simeq 0.5$ for the string+HDM model.
The parameter $S$ for the ICDM model is larger than the observed
constraint $S<0.52$ (95\% CL) given in Feldman et al. (2000).
A simple improvement is to increase the number $N$ of scalar
fields. This number $N$
corresponds to the degree of freedom in $\chi^2_N$-model. As discussed in
appendix A, the parameter $S$ scales are $S\propto N^{-1/2}$ and the
required number of freedom $N$ is $N\gsim (2.2/0.52)^2\simeq 20$
(Feldman et al. 2000).

In figure 3.a we show the nonlinear spectrum at the wave number
$k=k_{NL}$ as a function of the parameter $S$.  
The concentration parameter is fixed at $c(m)=9(m/m_*)^{-0.13}$ (Bullock
et al. 1999).
 We first calculate the ratio
\beq
\frac{ P_{NL}(k)}{P_L(k)}\equiv \frac{P_{2h}(k)+P_{1h}(k)}{P_L(k)},\lab{fullr}
\eeq
in the full dark-matter-halo approach.  We also evaluate the simple
approximation with putting $u(k,M)=1$, namely
\beq
\frac{P_{NL}(k)}{P_L(k)}\simeq 1+\frac{P_{1h}(0)}{P_L(k)}\lab{simap}
\eeq
to check the effects of the bias parameter $b(M)$ and the halo inner
density profile $u(k,M)$. As commented before,  validity of these two
elements for non-Gaussian models have not been clarified numerically so
far.  The 
simple approximation (\ref{simap})  shows reasonable agreement with the
full result (\ref{fullr}) and we
can expect that our analysis 
is not seriously affected by uncertainty of two elements in the
framework of the dark-matter-halo approach.  Figure 3.a
shows that nonlinear  spectrum  depends strongly on the
primordial non-Gaussianity measured by the parameter $S$.  For the
Gaussian model ($S=0$) we have  ${P_{NL}/P_L}\sim 1.3$ but
${P_{NL}/P_L}\sim 2$ for $S\simeq1$.  For the value
 $S=0.52$ (on 95\% CL of Feldman et al. 2000)   the correction becomes
${P_{NL}/P_L}\lsim 
1.55$ and  about 20\% larger than the Gaussian model. We also calculate
the ratio $P_{NL}/P_L$ for the Gaussian model using the fitting formula
of Peacock \& Dodds 
(1996).    The result is $\sim 1.2$  and close to our result $1.3$.

In figure 3.b we show the nonlinear spectrum  at the
wavenumber 
$k=0.66k_{NL}$ where the linear CDM spectrum becomes
$P_L(0.66k_{NL})=2P_L(k_{NL})$.  Apparently the correction becomes
smaller than  at 
$k=k_{NL}$.  We can easily understand this fact.  At weakly nonlinear
scale  we have 
\beq
 P_{NL}(k)-P_L(k)\simeq P_{1h}(0)
\eeq
and the nonlinear correction behaves as $P_{NL}/P_L\simeq
1+P_{1h}(0)/P_L(k)$ in the dark-matter-halo approach.

\section{SUMMARY}
In this article we have studied evolution of the matter power spectrum in
non-Gaussian models of structure formation. We have used the
dark-matter-halo approach that is an excellent method for analyzing
matter clustering from linear to nonlinear scales (Seljak 2000, Ma \&
Fry 2000b). This approach contains three basic ingredients (i) the mass
function $N(m)$ of dark-matter-halos, (ii) the bias factor $b(m)$ of
halos relative to the 
large-scale density fluctuations and (iii) the density profiles $u(k,m)$ 
of
halos.  Recent numerical simulations (Robinson \& Baker 2000) have
suggested 
 that  a simple extension of the
 Press \& Schechter formula provides  good fits to mass
function of clusters also in non-Gaussian models.
For the bias factor $b(m)$ and the density profiles $u(k,m)$ 
it has not checked
numerically whether we can simply extend the formulas of Gaussian models
to  non-Gaussian models. Therefore we take a moderate position
to limit our investigation up to weakly nonlinear regimes $k\lsim
k_{NL}$.  In these regions our analysis supports that the evolved power
spectrum does not depend on details of the bias factor $b(m)$ or the
density profile $u(k,m)$.   

The nonlinear power spectrum predicted by the dark-matter-halo approach
is constituted by two terms. The two-halo term counts two points in
different halos and the one-halo term represents contributions of  two
points within 
same halos.  We have shown that  the two-halo
term is almost same as the linear power spectrum in weakly nonlinear
regimes and nonlinear 
correction  mainly comes from  the one-halo term. This term is roughly
approximated by the integral of mass function $N(m)$ as $\int
N(m)m^2dm$.  Using this integral we can understand various aspects of
the 
weakly nonlinear effect. For example, nonlinear correction is larger
than a PDF with more positive tail or smaller spectral index $n$.

We also study a  COBE-normalized CDM spectrum in a flat model
($\Omega_0=0.3, \lambda_0=0.7$ and $h=0.7$) with log-normal PDFs. This
shape of PDF is motivated by work of Robinson \& Baker (2000) who found
numerically that PDFs of many representative non-Gaussian models are
fitted well by log-normal distributions.  We show that the nonlinear
 spectrum  $P_{NL}$ at $k=k_{NL}\simeq 0.2 {\rm
Mpc}^{-1}$ becomes  $P_{NL}/P_L=1.3$ for the Gaussian model but $\simeq
1.55$ for the 
log-normal model with $S=0.52$ that is on 95\% CL of the
 observed constraint
 by  Feldman et al. (2000).   
These results suggest that the ratios  $P_L/P_{NL}$ for $\Lambda$CDM model
at $k<k_{NL}\simeq 0.2{\rm Mpc}^{-1}$ could not  largely  (say
factor of 2) vary  within  currently allowed region of the primordial
non-Gaussianity.

\acknowledgements
The author would like to thank an anonymous referee for valuable comments
to improve this manuscript.
My work is  supported by Japanese  Grant-in-Aid No. 0001416.

\appendix

\section{SCALE DEPENDENCE OF THE SKEWNESS IN $\chi^2_N$-MODELS}
The scale dependence of a non-Gaussian PDF
becomes 
highly complicated as it is generally very difficult to deal with
infinite degrees of freedom  of 
 nonlocal  quantities.
Here we   study the  scale dependence of the PDF using the skewness
parameter  for
 $\chi^2_N$ model.

Let us discuss  $\chi^2_N$ distribution with $N$ degrees of freedom.
The unsmoothed density field $\rho(\vex)$ is written by $N$ independent
Gaussian random fields  $\phi_i(\vex)$ ($i=1,\cdots,N$)
\beq
\rho(\vex)=\sum_{i=1}^N \phi_i(\vex)^2.
\eeq
We assume that these $N$ fields have same statistical character. 
Moments of the unsmoothed field $\rho(\vex)$ is given in  terms  of the variance
$\lla\phi_i^2\rra$  of the basic Gaussian fields $\phi_i$ (the
bracket $\lla\cdot\rra$ represents ensemble average). We can easily
calculate their   moments as
\beqa
\lla \rho(\vex)\rra&=&N\lla \phi_i^2\rra,~~
\lla \delta\rho(\vex)^2 \rra=2N\lla \phi_i^2\rra^2,\nn\\ 
\lla \delta\rho(\vex)^3 \rra&=&8N\lla \phi_i^2\rra^3,~~ 
\lla \delta\rho(\vex)^4 \rra=(48N+12N^2)\lla \phi_i^2\rra^4.\lab{mom} 
\eeqa
where we have defined the overdensity $\delta\rho(\vex)$ by
$\delta\rho(\vex)\equiv  \rho(\vex)-\lla\rho(\vex)\rra$.
Furthermore we can derive the explicit formula for the  PDF of
$\rho $ as
\beq
p(\nu)d\nu=\frac{(1+\nu \sqrt{2N^{-1}})^{N/2-1}}{(2N^{-1})^{(N-1)/2}
\Gamma(N/2)} \exp\lmk-\frac{N}2 \lmk1+\sqrt{\frac2N}\nu \rmk  \rmk
d\nu, \lab{pdf}
\eeq
where we have defined the normalized overdensity $\nu$ by $\nu\equiv
\delta\rho/\lla\delta\rho^2\rra^{1/2}$ and $\Gamma(\cdot)$ is the  Gamma
function.  
As the number $N$ increases, the
 PDF becomes closer to the Gaussian distribution. This seems
reasonable considering the central limit theorem.  For smaller degrees
of freedom  $N$,  the
 PDFs show strong non-Gaussianity.

 The skewness parameter $S$
is a fundamental quantity to characterize the 
 non-Gaussian profile.  For
$\chi^2_N$-model we have the  following expression from equation
(\ref{mom})  for moments
\beq
S\equiv \frac{\lla \delta\rho(\vex)^3 \rra}{{\lla \delta\rho(\vex)^2
\rra}^{3/2}}= \sqrt{\frac8N}. \lab{skew}
\eeq
Note that using the skewness parameter $S$,  we can
specify  the
$\chi_N^2$-distribution  that is completely characterized by the degrees of
freedom  $N$.

The simplicities of  relations given so far are mainly due  to  local nature of
the  density field
$\rho(\vex)$. Every statistical characters of the field  are reduced
to  simple Gaussian statistics of the basic fields $\phi_i(\vex)$.  When
we discuss 
smoothing effects on the density field $\rho(\vex)$ that is an
nonlinear combination of Gaussian fields $\phi_i$, the situation becomes
highly 
complicated. 
First we define the smoothed density field $\rho_R(\vex)$ by
\beq
\rho_R(\vex)=\int d\vex' \rho(\vex') w(\vex'-\vex:R),
\eeq
where $w(\vex;R)$ is a filter function with smoothing  radius $R$.  In
this 
appendix we mainly use the Gaussian filter 
\[
 w(\vex:R)=\frac1{\sqrt{(2\pi R^2)^3}}\exp\lmk-\frac{\vex^2}{2R^2} \rmk.
\]
  We expand the fields $\phi_i(\vex)$
 in terms of the Fourier modes $\phi_i(\vek)$ 
\beq 
\phi_i(\vex)=\frac1{(2\pi)^3}\int d\vek \phi_i(\vek)\exp(i\vek\cdot \vex).
\eeq 
 As the basic fields $\phi_i$ obey  independent random-Gaussian
distributions,
their statistical nature is completely determined by their  power spectrum
$P_\phi(k)$ defined by 
\beq
\lla \phi_i(\vek)\phi_j (\vel)\rra=P_\phi(k)\delta_{ij}\delta_D(\vek+\vel),
\eeq
where $\delta_{ij}$ is the  Kronecker's delta,  $\delta_D(\cdot)$ is the
Dirac's delta function and we have assumed that fluctuations are isotropic.
Similarly the smoothed density field $\rho_R(\vex)$ is written by 
 $\phi_i(\vek)$  as
\beq
\rho_R(\vex)=\sum_{i=1}^N \frac1{(2\pi)^6} \int d\vek d\vel
\phi_i(\vek)\phi_i (\vel) \exp(i(\vek+\vel)\cdot\vex)W(|\vek+\vel|R),\lab{fou}
\eeq
where $W(kR)$ is the Fourier transformed filter function of
$w(\vex:R)$. For the  
 Gaussian filter 
we have $W(kR)=\exp(-k^2R^2/2)$. 
With expression (\ref{fou}) we can write down the second- and third-order moments
for the smoothed density field in terms of the  power spectrum
$P_\phi(k)$. After some algebra we arrive at 
\beqa
\lla \delta \rho_R^2\rra &=&\frac{2N}{(2\pi)^6} \int d\vek d \vel P_\phi(k)
P_\phi(l) W(|\vek+\vel|R)^2\lab{2nd},\\
\lla \delta \rho_R^3\rra &=&\frac{8N}{(2\pi)^9} \int d\vek d \vel d \vem
P_\phi(k) P_\phi(l)  P_\phi(m) W(|\vek+\vel|R)  W(|\vel+\vem|R)  W(|\vem+\vek|R). \lab{3rd}
\eeqa
Using these  equations  we evaluate the skewness parameter $S_R$  for the
smoothed 
density field.  We can easily confirm  that the unsmoothed value
$S=\sqrt{8/N}$ is recovered by setting $R=0$ in these equations.
The smoothing effect changes  the skewness parameter
$S_R$  from the unsmoothed value $S=\sqrt{8/N}$.
  To discuss quantitative effects of smoothing we define a parameter
$f_R$ defined by
\beq
S_R\equiv \frac{\lla \delta\rho_R(\vex)^3 \rra}{{\lla \delta\rho_R(\vex)^2
\rra}^{3/2}}= \sqrt{\frac8{f_RN}}.
\eeq
Note that the parameter $f_R$ does not depend on the degrees of freedom $N$. 
  The effective
degrees of freedom  $N'$  determined form the smoothed skewness
parameter 
$S_R$ becomes $f_R $ times original value.

Next let us  evaluate the smoothed skewness $S_R$ numerically and
the 
scaling parameter $f_R$ for a sequence of matter fluctuations.  We adopt
the following from  for the
 power spectrum of the Gaussian fields
$\phi_i$
\beq
P_\phi(k)=A k^{n_\phi} \exp(-k^2 r^2).\lab{ps}
\eeq
As the normalization factor $A$ is irrelevant for our linear analysis we
simply put $A=1$.  Due to the Gaussian cut-off $\exp(-k^2r^2)$,  
small scale power at wave-number 
$k\gsim r^{-1}$ is strongly suppressed and the fields $\phi_i(\vex)$
and $\rho(\vex)$ are very smoothed at spatial scale smaller than $ r $.  
This means that the smoothed density field $\rho_R(\vex)$ is  close to the
unsmoothed field $\rho(\vex)$ for smoothing radius $R$ with 
$R\lsim r$.  Therefore  radius $r$ represents
the  
spatial scale where the PDF of the smoothed field $\rho_R$ would become
close to the original $\chi^2_N$ 
distribution and the scaling factor would be $f_R\sim 1$.

Using equation (\ref{2nd})
 for  variance of the  smoothed density field, we obtain the
 following result for the Gaussian filter
\beq
\lla \delta \rho_R^2\rra =
\frac{N R^{-6-2n_\phi}}{4\pi^4}  \int_0^\infty  ds \int_0^\infty dt
(st)^{n_\phi+1} 
  \exp\lkk-\lmk1+\frac{r^2}{R^2} \rmk (s^2+t^2)\rkk \sinh
(2 s t ). \lab{2nd2} 
\eeq
At large smoothing  radius $R$ we have $\lla (\delta\rho)^2\rra\propto
R^{-6-2n_\phi}$ and the matter power spectrum at small wave-number 
would behave as (Peebles 1999b, White 1999, Scoccimarro 2000)
\beq
P_\rho(k)\propto k^{2n_\phi+3}=k^{n_\rho},\lab{nphi}
\eeq
where we have defined the spectral index $n_\rho$ of density field by
$n_\rho\equiv 2n_\phi+3$.

The expression (\ref{3rd}) for the third-order moment is a
nine-dimensional 
integral. But using symmetry with respect to variables $\vek,\vel$ and $\vem$,
we can simplify the expression. After some   algebra we arrive at
\beqa
\lla \delta \rho_R^3\rra&=&\frac{N R^{-9-3n_\phi}}{4\pi^6} \int_0^\infty  ds \int_0^\infty dt \int_0^\infty du \int_0^\pi da \sin a  \int_0^\pi db \sin b  (s t
u)^{2+n_\phi} I_0( t u \sin
a \sin b)\nn\\
& & \times \exp\lkk-\lmk1+\frac{r^2}{R^2} \rmk (s^2+t^2+u^2)- \lmk s t\cos a+tu
\cos a \cos b +us \cos b \rmk   \rkk ,\lab{3rd2}
\eeqa  
where $I_0(x)$ is the $0$-th modified Bessel function.    
From equations (\ref{2nd2}) and (\ref{3rd2}), it is apparent that the  factor $f_R$ is
written in the  form  
\beq
f_R=f(R/r).
\eeq

In the large scale limit $R\to \infty$, the small scale Gaussian
 cut-off  in
the power spectrum $P_\phi(k)$ (eq.[\ref{ps}]) would have 
 no effects on the second- and third-order  moments as expected from
equations (\ref{2nd2}) and (\ref{3rd2}). In this limit we can put $r=0$
and our results should coincide with results for
 pure power-law models $P_\phi(k)\propto
k^{n_\phi}$ (Peebles 1999b, Scoccimarro 2000). Simple analytic expressions for the matter
power spectrum 
$P_\rho$ and bispectrum $B_\rho$ for $n_\phi=-2$ model are given by
Scoccimarro (2000) as follows
\beq
P_{\rho}(k)=\frac{2\pi^3N}k, ~~~
B_{\rho} (\vek,\vel,\vem)=\frac{8\pi^3N}{klm}.
\eeq
With the latter expression we can  numerically study 
 the third-order moment $\lla( \delta\rho_R)^3\rra$ at $R\to\infty$.
In this case we need only two-dimensional numerical integration that
is  much simpler than  the five-dimensional
integration (\ref{3rd2}).  
 We find the numerical result  \footnote{For the top-hat filter we have
similar result $f(\infty)=2.6$.}  
\beq
f(\infty)=2.6~~~(n_\phi=-2).\lab{mn2}
\eeq
This asymptotic value can be used to check our five-dimensional
numerical integration required to evaluate the smoothed skewness at an
intermediate scale.

In figure 4 we show  scale dependence of the factor $f(R/r)$ for
power spectrum (\ref{ps}) with various spectrum indices, 
$n_\phi= -1$, $n_\phi= -1.5$, $n_\phi= -2$
and $n_\phi= -2.4$. These indices  correspond to $n_\rho=1$,  $n_\rho=0$,
 $n_\rho=-1$, and  $n_\rho=-1.8$, respectively. The last choice  $n_\phi=
-2.4$ is same as  Peebles (1999b).
As expected,  we have $f(R/r)\simeq 1$ for $R\lsim r$ reflecting the 
original profile of the  unsmoothed PDF.    For  a larger smoothing
radius  $R\gsim 20 r$ the parameter
$f(R/r)$ relaxes to 
an asymptotic value  $f(\infty)$ that 
 does  not depend on the small scale cut-off as explained earlier.
 The asymptotic value for $n_\phi=-2$ model is close to the value given
in equation (\ref{mn2}) and show the validity of our numerical integration.
Peebles (1999b) calculated the skewness parameter  smoothed by
the top-hat filter for a model with one degree of freedom  $N=1$ and
spectral index at 
$n_\phi=-2.4$, using Monte Carlo 
integration. 
His result is $S_R=2.46$ and corresponds to $f_R\simeq 1.32$. This value
is close to  our asymptotic value $f(\infty)\simeq 1.37$ obtained for the
Gaussian filter.

As shown in figure 4, the factor $f(R/r)$ largely depends on the
spectral indices $n_\phi$. We have $f(\infty)\simeq 200$ for
$n_\phi=-1 ~(n_\rho=1)$ model. This large value indicates that the PDF
would become close to the  Gaussian profile. The factor keeps
$f(R/r)\lsim 3$ for $n_\phi=-2~ (n_\rho=-1)$ model. White (1999)
investigated this model using N-body simulations and found that the PDF
is skewed for Gaussian smoothing spanning a factor 5 in scale. This
behavior  is
consistent with our result.  For $n_\phi=-2.4$ model, the factor
$f(R/r)$ changes only $\sim 40\%$ in the range $0\le R <\infty$. These
 dependence on the spectral indices
$n_\phi $
seems reasonable considering that as $n_\phi$ increases,
number of statistically independent region would also increases for a
given 
smoothing volume and the PDF would become a more Gaussian like profile.

\newpage

\begin{figure}
\epsscale{0.4}
\plotone{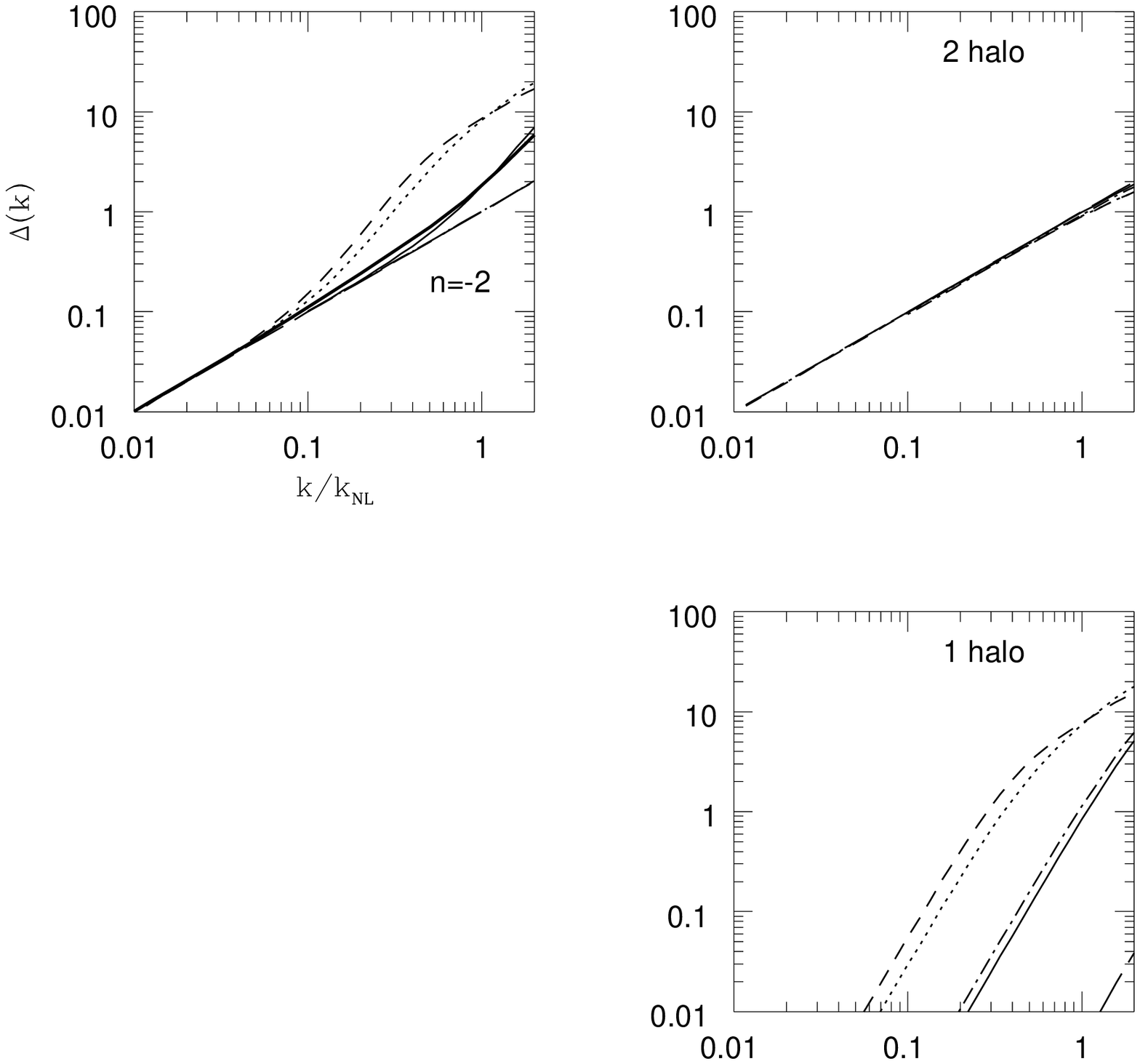}
\plotone{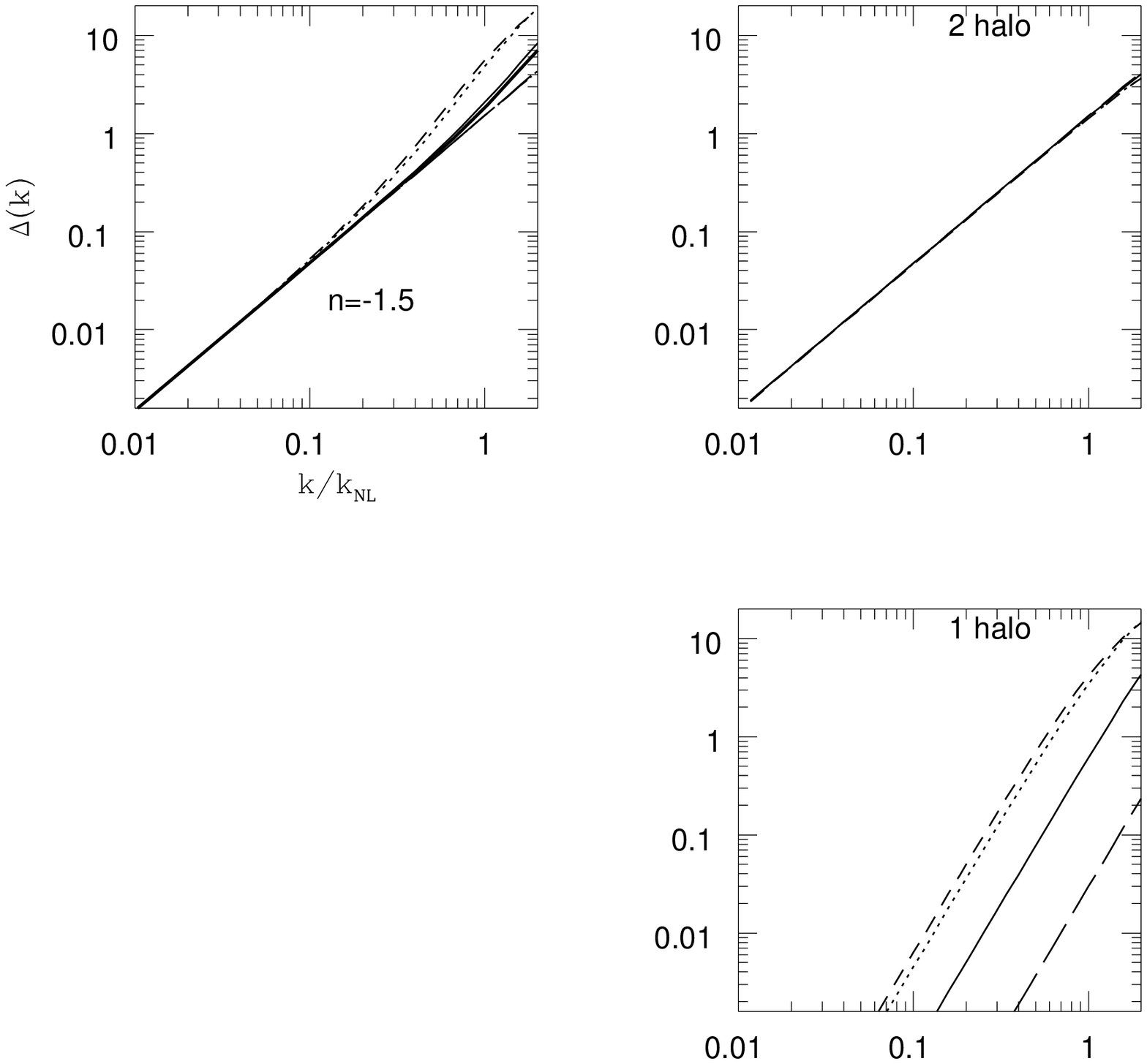}
\plotone{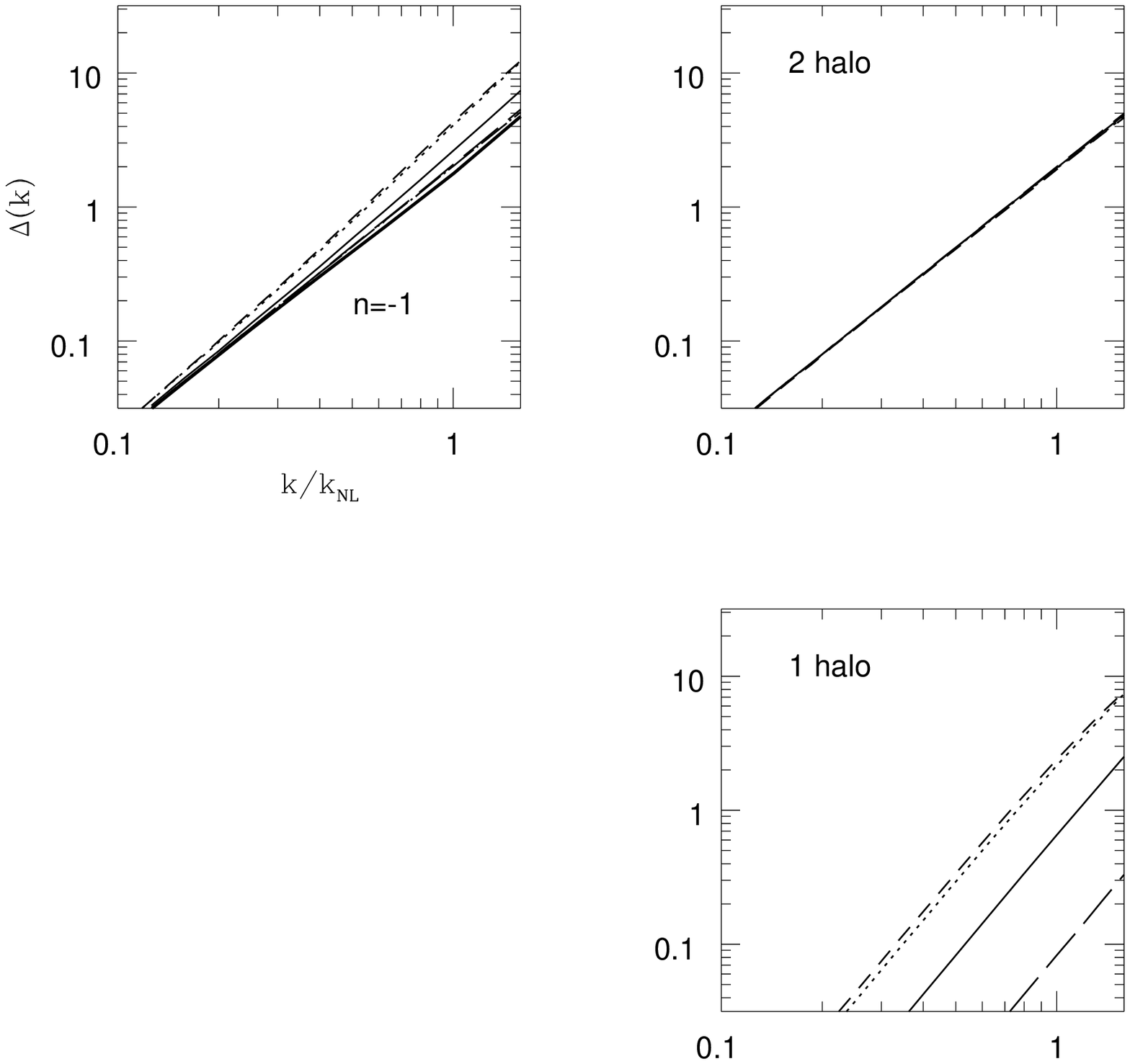}
\caption{Nonlinear power spectrum 
	   $\Delta(k)\equiv P_{NL}(k)k^3/(2\pi^2)$ for $n=-2$ model
	   (fig.1a),  
	   $n=-1.5$ model (fig.1b), $n=-1$ model (fig.1c). 
	   In the right panels we plot the two-halo and one-halo terms
	   separately.  
	   The total values (summations of two terms) are presented in
	   the left  
	   panels. 
	   Dash-dotted lines represent the linear spectra.   
	   Thin solid lines correspond to the G-model, dotted-lines to
	   the C-model, 
	   short-dashed lines to the LN1-model and long-dashed lines to
	   the 
	   LN2-model. The thick-solid lines are obtained from the fitting
	   formula of Peacock \& Dodds (1996) given for  Gaussian
	   fluctuations.  In the left panels the long-dashed lines
 (LN-2 model) and the dash-dotted lines (linear spectra) are nearly
 overlapped.}
\end{figure}

\begin{figure}
\plotone{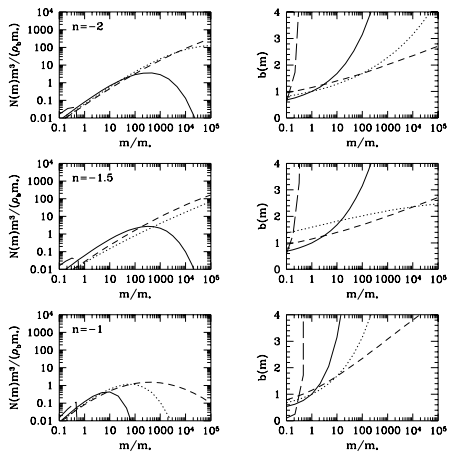}
\caption{The mass functions and  the bias parameters for various PDFs.
	   Correspondence of lines and models is same as  figure
	   1. There are no halos with $m\ge m_*$ in the LN2-model. We
	   plot the 
	   mass function $N(m)$ in the form $N(m)m^3$, as the integral
	   $\int N(m)m^2dm$ is important for our analysis. }
\end{figure}

\begin{figure}
\plotone{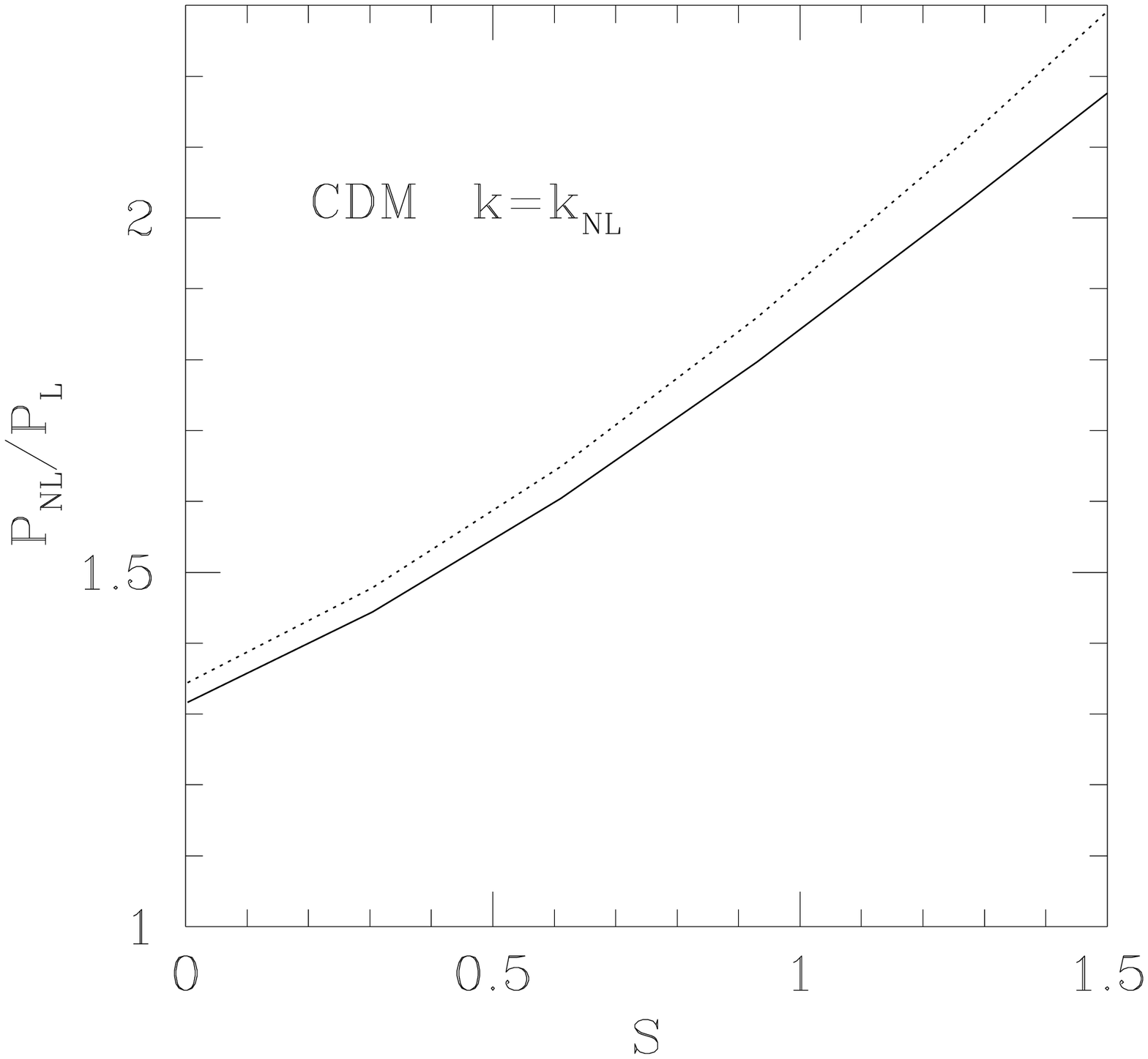}
\plotone{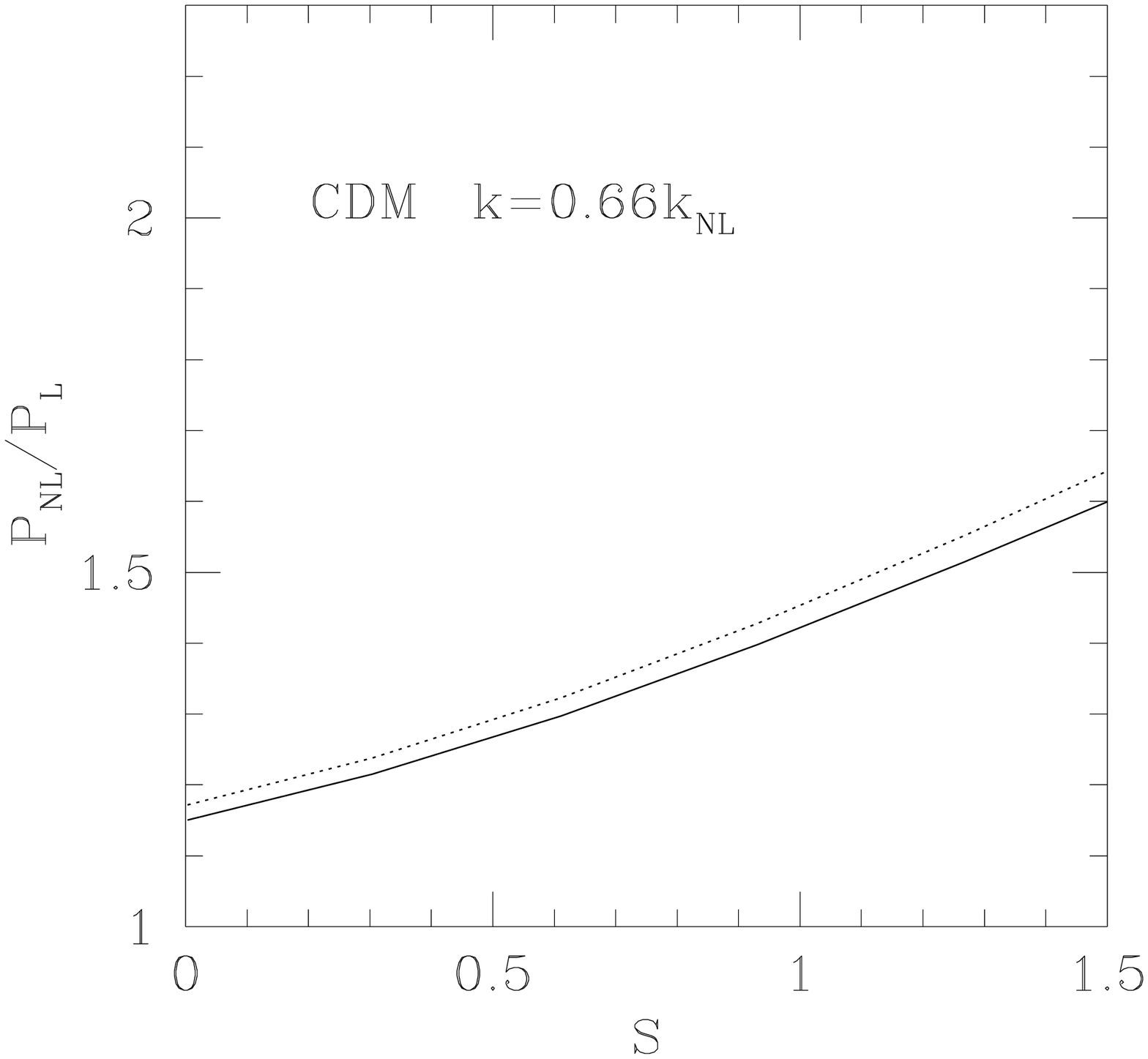}
\caption{Nonlinear  spectrum for non-Gaussian
 initial fluctuations with CDM spectra.  We characterize the
 non-Gaussianity  by the skewness parameter $S$. The ratio
$P_{NL}/P_L$ is given  at $k=k_{NL}$ (fig 3.a) and $k=0.66k_{NL}$ (fig
 3.b). The solid lines represent the results from full dark-matter-halo
 approach and the dotted lines are obtained by the simple approximation
 described in the main text.}
\end{figure}

\begin{figure}
\epsscale{0.5}
\plotone{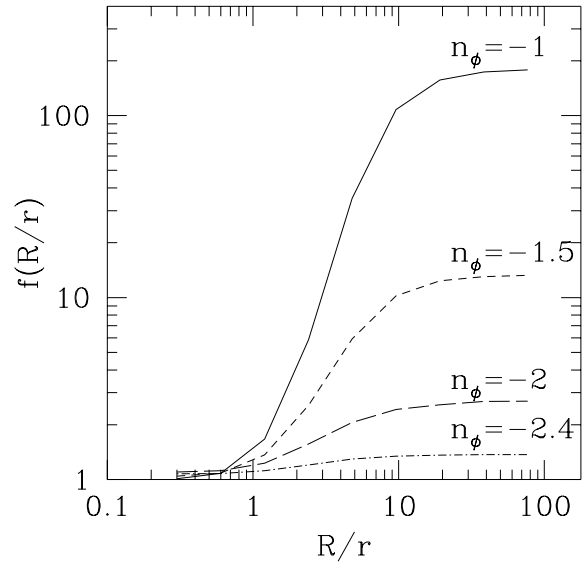}
\caption{The scaling  factor $f(R/r)$ (see eq.[A11]) as a function of Gaussian smoothing
         radius $R$. We have $f(R/r)\sim 1$  for $R/r > 1$ and
         $f(R/r)\sim const $ for  $R/r> 20$.}
\end{figure}

\end{document}